\documentclass[prb,reprint,superscriptaddress,longbibliography]{revtex4-1}
\usepackage{graphicx}
\usepackage[caption=false]{subfig}
\usepackage{bm}
\usepackage{amsmath}
\usepackage{amssymb}
\usepackage{mathtools}
\usepackage{siunitx}
\usepackage{multirow}
\usepackage{array}
\usepackage{xcolor}
\usepackage{longtable}
\usepackage[version=4]{mhchem}
\usepackage{hyperref}
\hypersetup{colorlinks=true,
            linkcolor=red,
            citecolor=blue,
            urlcolor=green}
 
\newcolumntype{K}[1]{>{\centering\arraybackslash}p{#1}}

\AtBeginDocument{\usepackage{booktabs}}

\begin{document}
\title{Two-dimensional Boron Monosulfides: Semiconducting and Metallic Polymorphs}

\author{Dong Fan} 
\affiliation{College of Materials Science and Engineering, Zhejiang University of Technology,Hangzhou 310014, China} 

\author{Chuncheng Yang} 
\affiliation{College of Materials Science and Engineering, Zhejiang University of Technology,Hangzhou 310014, China} 

\author{Shaohua Lu}
\email{lsh@zjut.edu.cn}
\affiliation{College of Materials Science and Engineering, Zhejiang University of Technology,Hangzhou 310014, China} 

\author{Xiaojun Hu}
\email{huxj@zjut.edu.cn}
\affiliation{College of Materials Science and Engineering, Zhejiang University of Technology,Hangzhou 310014, China}

\begin{abstract}
The typical two-dimensional semiconductors, group \uppercase\expandafter{\romannumeral3A} chalcogenides, have garnered tremendous interest for their outstanding electronic, mechanical, and chemical properties. However, so far, there have been almost no reports on boron monosulfides (BS) binary material. Here, four two-dimensional BS sheets, namely the $\alpha$-, $\beta$-, $\gamma$-, and $\delta$-BS sheets, are proposed and discussed from $\emph{ab initio}$ calculations. State-of-the-art first principles calculations reveal all these structures are thermally and dynamically stable, indicating the potential for successful experimental synthesis. Especially, for $\alpha$-BS, it has a calculated exfoliation energy of 0.96 J m$^{-2}$, suggesting the preparation of  $\alpha$-BS is feasible by the exfoliation of bulk rhombohedral-BS. Our results show that $\alpha$-, $\beta$-, $\gamma$-BS are semiconductors, whereas $\delta$-BS is a metallic system. Remarkably, our calculations indicate that $\delta$-BS is a superconductor with a large electron-phonon coupling ($\lambda$ = 1.51) leading a high superconducting critical temperature ($T_c$ $\approx $ 21.56 K), which is the first report of intrinsic superconducting property among all two-dimensional group \uppercase\expandafter{\romannumeral3A} chalcogenides. The desired mechanical and electronic properties render the BS sheets as the promising two-dimensional materials for future applications in nanoelectronics.

\end{abstract}


\maketitle

\section{INTRODUCTION}

In recent years, there has been a rapidly increasing research interest in two-dimensional (2D) layered materials, since the physical properties of 2D crystals often remarkably change from those of their bulk counterparts, providing a new alternative for applications in nanodevices. The most representative sample is graphene, because of its excellent electronic, mechanical, and thermal properties.\cite{novoselov2004electric,neto2009electronic} Inspired by pioneering work of Novoselov and Geim $\emph{et al.}$,\cite{novoselov2004electric} the subject of discovering new 2D materials other than graphene is one of the most significant fields of current scientific research. Beyond graphene, there is a very wide spectrum of 2D crystals that range from insulators to semiconductors and metals.\cite{butler2013progress,bhimanapati2015recent} $\emph{i.e.}$, hexagonal boron nitride ($h$-BN),\cite{pakdel2014nano} transition metal chalcogenides (TMCs),\cite{heine2014transition}, metal carbides and nitrides (MXenes),\cite{anasori20172d} and  mono-elemental 2D materials.\cite{mannix2017synthesis} However, compared to bulk materials, there is still a lack of variety and assortment of 2D materials.

Another new family of 2D crystals, 2D group \uppercase\expandafter{\romannumeral3A} chalcogenides, are promising materials for photoelectronics,\cite{hu2013highly} gas sensing,\cite{yang2014high} and Li-ion Battery anodes\cite{zhang2017enabling}. Until now, various layered group \uppercase\expandafter{\romannumeral3A} chalcogenides, $\emph{i.e.}$, GaS,\cite{hu2013highly} GaSe,\cite{lei2013synthesis,jie2015layer}  GaTe,\cite{wang2014role,wang2015high} and InSe,\cite{mudd2013tuning} have been  synthesized experimentally. Following by these achievements, the great endeavors have also been done to investigate the intriguing physical and chemical properties of these materials for potential applications in many fields.\cite{hu2013highly,yang2014high,zhang2017enabling,xu2016synthesis} In fact, bulk boron-sulfur binary material, is also a light-element member of group \uppercase\expandafter{\romannumeral3A} chalcogenides family. Experimentally, boron-sulfur binary compounds have been known since 1977,\cite{diercks1977crystal,krebs1980b8s16} and its monosulfides structure were reported in 2001,\cite{sasaki2001high} bulk rhombohedral boron monosulfide ($r$-BS), namely. However, unlike other common group \uppercase\expandafter{\romannumeral3A} chalcogenides semiconductors, little research has been done toward preparation of the 2D binary BS compounds due to the lack of knowledge of their physical properties.\cite{demirci2017structural} To date, only several structural and electronic properties of bulk binary BS compounds have been reported, showing that the bulk $r$-BS is a semiconductor with an estimated band gap of 3.4 eV.\cite{sasaki2001high,demirci2017structural}

In this letter, we present four new BS monolayer structures, named $\alpha$-, $\beta$-, $\gamma$-, and $\delta$-BS, predicted $\emph{via}$ combined $\emph{ab initio}$ calculations  and structure search method. Both dynamical and thermal stability of these sheets is investigated by phonon spectrum calculation and $\emph{ab initio}$ molecular dynamics (AIMD) simulation. The electronic structure calculations show that the $\alpha$-, $\beta$-, and $\gamma$-BS are semiconductors with the band gaps of 4.03, 3.89, and 2.94 eV, respectively; whereas $\delta$-BS is a metallic sheet. More importantly, electron-phonon coupling calculations show that $\delta$-BS is superconducting with a high superconducting critical temperature ($T_c$) of 21.56 K.

\section{COMPUTATIONAL METHODOLOGY}
The search of stable BS systems is performed using the structure particle swarm optimization (PSO) algorithms as implemented in the CALYPSO package.\cite{wang2014perspective,wang2010crystal} All the calculations were carried out with the Vienna $\emph{Ab initio}$ Simulation Package (VASP).\cite{kresse1993ab} The density functional theory (DFT) with the generalized gradient approximation (GGA) of Perdew-Burke-Ernzerhof (PBE) functional were employed.\cite{blochl1994projector,perdew1996generalized}  The kinetic cutoff of the plane wave was set to 650 eV and the precision of energy convergence was 10$^{-5}$ eV. The atomic positions were relaxed until the maximum force on each atom was less than 10$^{-3}$ eV$/$\r{A}. All 2D structures were simulated with a periodic boundary condition, and the vacuum region between adjacent layers was fixed to at least 20 \r{A}  in order to eliminate possible interaction. Electronic band structures were obtained by he Heyd-Scuseria-Ernzerhof (HSE06) screened hybrid functional\cite{paier2006screened} and compared with those with conventional PBE functional. Phonon dispersions and frequency densities of states (DOS) were performed in Phonopy code\cite{togo2015first} interfaced with the density functional perturbation theory (DFPT)\cite{baroni2001phonons} as carried out in VASP code. AIMD simulations are performed to determine the thermal stability of  sheets. The uniaxial strain was applied along the x and y directions to simulate the stress-strain relationship of sheets, defined as ${\epsilon}$ = (a - a$_0$)/a$_0$, where a and a$_0$ are the lattice constants with and without strain, respectively. The van der Waals (vdW) interaction is incorporated by the vdW-DF method proposed by Dion $\emph{et al}$.\cite{dion2004van} to estimate the exfoliation energy. The Quantum-ESPRESSO 6.1 package\cite{giannozzi2009quantum} is used to study the electron-phonon coupling (EPC) in $\delta$-BS. The norm-conserving pseudopotentials (FHI) with a cutoff energy of 120 Ry were used. The Gaussian smearing method with a smearing parameter of $\sigma$ = 0.02 Ry was used.
 
 \begin{figure}[htbp]
\centering
\includegraphics[width=0.9\columnwidth]{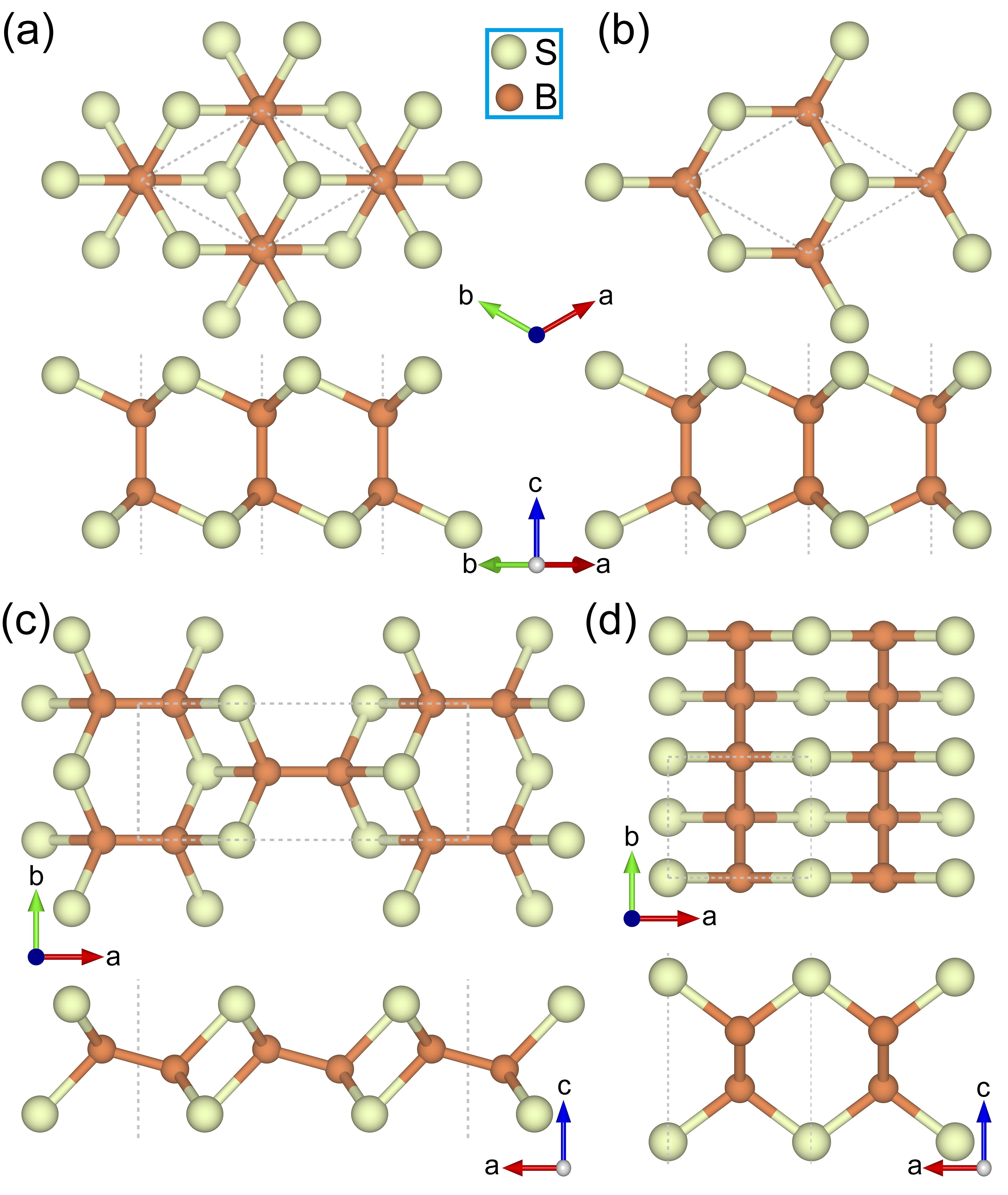}
\caption{Top (upper) and side (lower) views of the optimized geometric structure of (a) $\alpha$-, (b) $\beta$-, $\gamma$-, and $\delta$-BS. $a$ and $b$ represent the lattice vectors; the unit cell is indicated by dashed lines.}
\end{figure}

\section{RESULTS AND DISCUSSION}

$\textbf{Crystal structures.}$ Optimized structures of all predicated BS sheets are shown in Figure 1, and the calculated lattice parameters are summarized in Table S1. We identify four different phases, denoted by $\alpha$-, $\beta$-, $\gamma$-, and $\delta$-BS, as shown in Figures 1a-1d, respectively. Evidently, $\alpha$- and $\beta$-BS sheets share similar structural features with the previously fabricated 2D hexagonal GaS and GaSe.\cite{late2012gas,hu2012synthesis} They adopt the hexagonal lattices with two B and two S atoms in each unit cell. The optimized lattice constants for $\alpha$- and $\beta$-BS are $a$ = $b$ = 3.06, and $a$ = $b$ = 3.04 \r{A}, respectively. Also, the obtained results for $\beta$-BS is very well in good agreement with previous theoretical work.\cite{demirci2017structural} However, the $\gamma$-BS and $\delta$-BS crystallize in the orthorhombic lattice with space group $\emph{C2/m}$ and $\emph{Pmma}$, showing the $\emph{C$_{2h}$}$ and $\emph{D$_{2h}$}$ symmetry, respectively. As shown in Figures 1c-1d, 1 unit cell of $\gamma$-BS ($\delta$-BS) monolayer consists of 4 (2) B atoms and 4 (2) S atoms with the optimized lattice parameters being $a$ = 7.38 (3.07) \r{A} and $b$ = 3.06 (2.59) \r{A}, respectively. 

\begin{figure}[htbp]
\centering
\includegraphics[width=0.95\columnwidth]{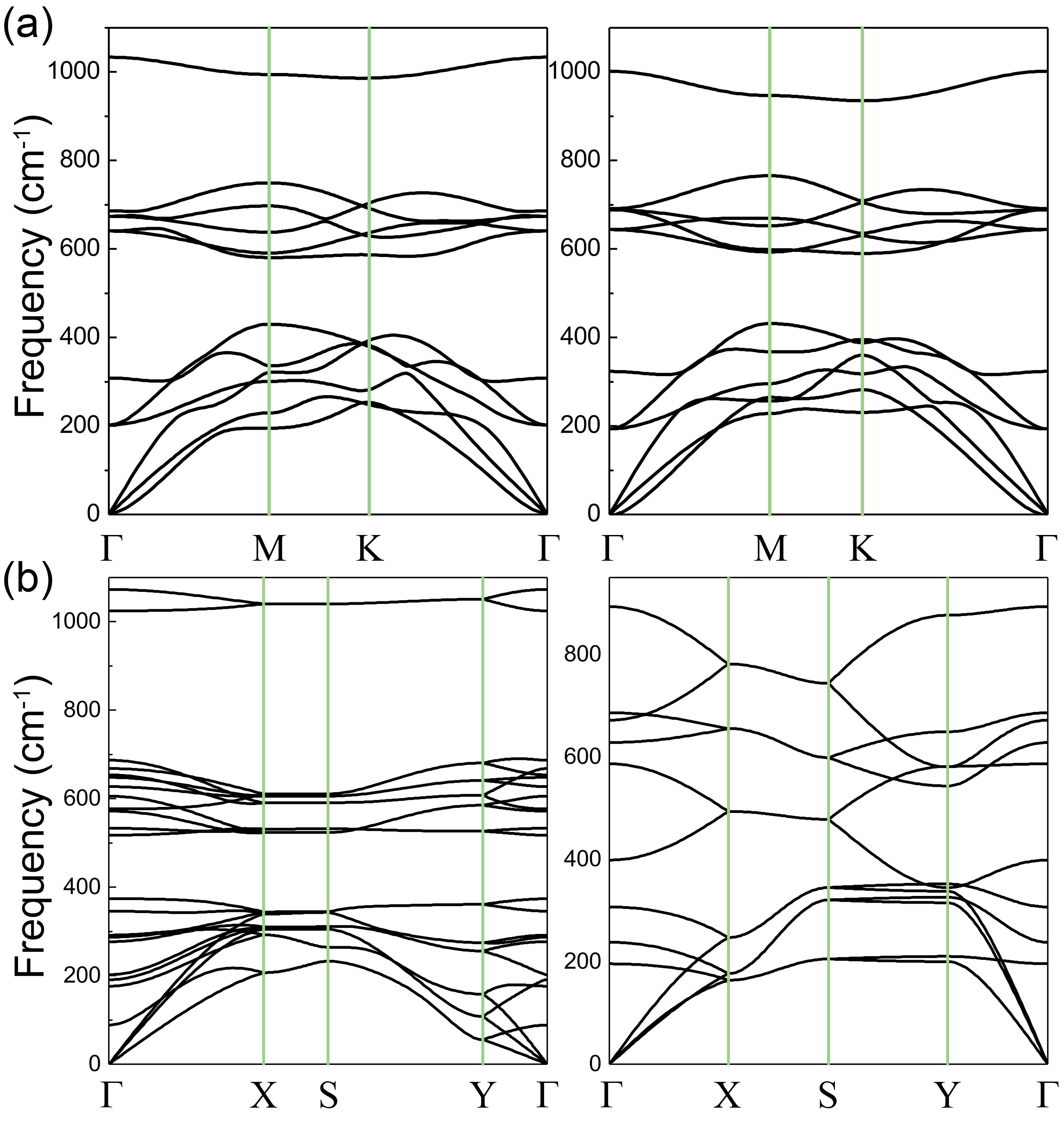}
\caption{Computed phonon spectrum of of (a) $\alpha$- (left), $\beta$- (right), (b) $\gamma$- (left), and $\delta$-BS (right).}
\end{figure}

$\textbf{Stabilities.}$ To examine the relative stability of these different allotropes, the cohesive energy (E$_c$) per atom with respect to the energy of an isolated B and S atoms is calculated, as defined as E$_{coh}$ = (nE$_B$ + mE$_S$ - E$_{BS}$)/${n+m}$, in which E$_B$, E$_S$, andE$_{BS}$ are the calculated total energies of a single B atom, a single S atom, and the BS sheet, respectively; $n$ ($m$) is the number of B(S) atoms in the unit cell. According to our calculations, these monolayers have the cohesive energies of 5.23, 5.22, 5.11 and 4.92 eV per atom for $\alpha$-, $\beta$-, $\gamma$-, and $\delta$-BS, respectively.  Thus, $\alpha$-BS is a energetically most stable phase, while $\beta$-, $\gamma$-, and $\delta$-BS are the metastable phases. As a reference, the cohesive energies of the experimentally realized 2D silicene, and phosphorene, are 3.71 and 3.61 eV per atom, respectively.\cite{wang2016semi} Therefore, the even higher cohesive energies can ensure that the proposed monolayers are strongly bonded with the unique networks.

The dynamical stability of these monolayers can be further checked by phonon dispersion curves, as shown in Figure 2. No imaginary phonon frequencies were observed in the whole Brillioun zone, suggesting their dynamical stability. The highest frequency of $\gamma$-BS reaches up to 1073 cm$^{-1}$ , higher than that of 473 cm$^{-1}$ in MoS$_2$,\cite{molina2011phonons} t-SiC (735 cm$^{-1}$),\cite{fan2017novel} and silicene (580 cm$^{-1}$),\cite{cahangirov2009two} indicating the strong B-S and B-B bonds in the structures. Additionally, their thermal stability is also confirmed by performing the AIMD simulations, as shown in the Figures S1-S2. Therefore, the above-mentioned results demonstrate that all these monolayers have satisfactory energetic, dynamical, and thermal stability.

\begin{figure}[htbp]
\centering
\includegraphics[width=0.99\columnwidth]{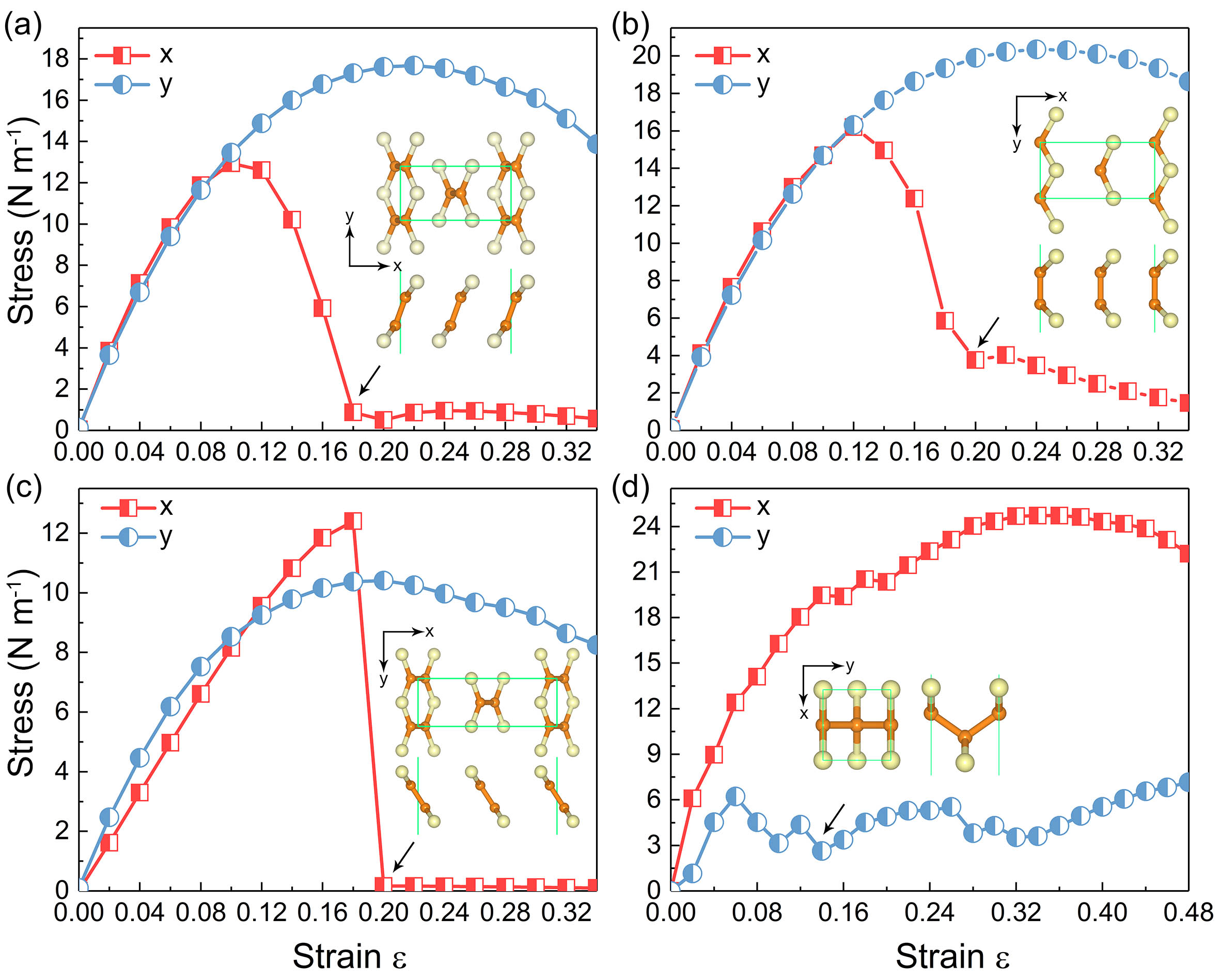}
\caption{The calculated strain-stress relation for (a) $\alpha$-, (b) $\beta$-, (c) $\gamma$-, and (d) $\delta$-BS structures. The insets illustrate the atomic structures of BS sheets at the structural failure point.}
\end{figure}

$\textbf{Mechanical Properties.}$ The mechanical properties are indispensable for new materials applications in the real world. Conventionally, for a mechanically stable 2D free-standing configuration, the calculated elastic constants should satisfy C$_{11}$C$_{22}$-C$_{12}$C$_{21}$ $>$ 0 and C$_{66}$ $>$ 0.\cite{wei2014superior} As listed in Table S1, all the calculated elastic constants of the proposed new structures satisfy the this criteria, indicating that these 2D compounds have favorable mechanical stability. As listed in Table S1, the in-plane Young's modulus (or  in-plane stiffness) is calculated to be 212 GPa$\cdot$nm for $\beta$-BS, which is distinctly higher than that of experimentally synthesized 2D GaS (73 GPa$\cdot$nm) and silicene (61 GPa$\cdot$nm).\cite{demirci2017structural,zhang2011stability} However, for $\gamma$- and $\delta$-BS sheets, as elastic constant C$_{11}$ is not equal to C$_{22}$, indicating they are mechanically anisotropic.

Besides in-plane Young's modulus, ideal strength is also an important mechanical property for 2D material. The ideal tensile stress versus strain for the BS sheets is shown in Figure 3. With small strains  deformations, the sheets exhibit linear stress-strain relationship (with distinguished elastic anisotropy for $\gamma$- and $\delta$-BS). As the applied strains increase, their stress-strain behaviors become nonlinear and show difference changing trends along the x and y directions. Particularly, for $\alpha$- and $\beta$-BS, along y direction, both the peak strengths and the corresponding critical strains are higher than along x direction, while the opposite trends occurred in $\gamma$- and $\delta$-BS sheets. The $\alpha$-BS can sustain stress up to 13 N m$^{-1}$ and 18 N m$^{-1}$ in the x and y directions, respectively. The corresponding critical trains are 0.1 (x) and 0.22 (y). The ideal strengths for the $\beta$-BS are 16 N m$^{-1}$ and 20 N m$^{-1}$ in the x and y directions, respectively, and their critical strains are 0.12  (x) and 0.24 (y). Figure 3d presents the results for $\delta$-BS, its peak strength is 25 N m$^{-1}$ at ${\epsilon}_x$  = 0.34 and  6 N m$^{-1}$ at ${\epsilon}_y$ = 0.06, respectively. Thus, the ideal strengths of BS sheets are significantly higher than other 2D materials, such as borophenes, MoS$_2$, and phosphorene.\cite{zhang2017elasticity}

\begin{figure}[htbp]
\centering
\includegraphics[width=0.9\columnwidth]{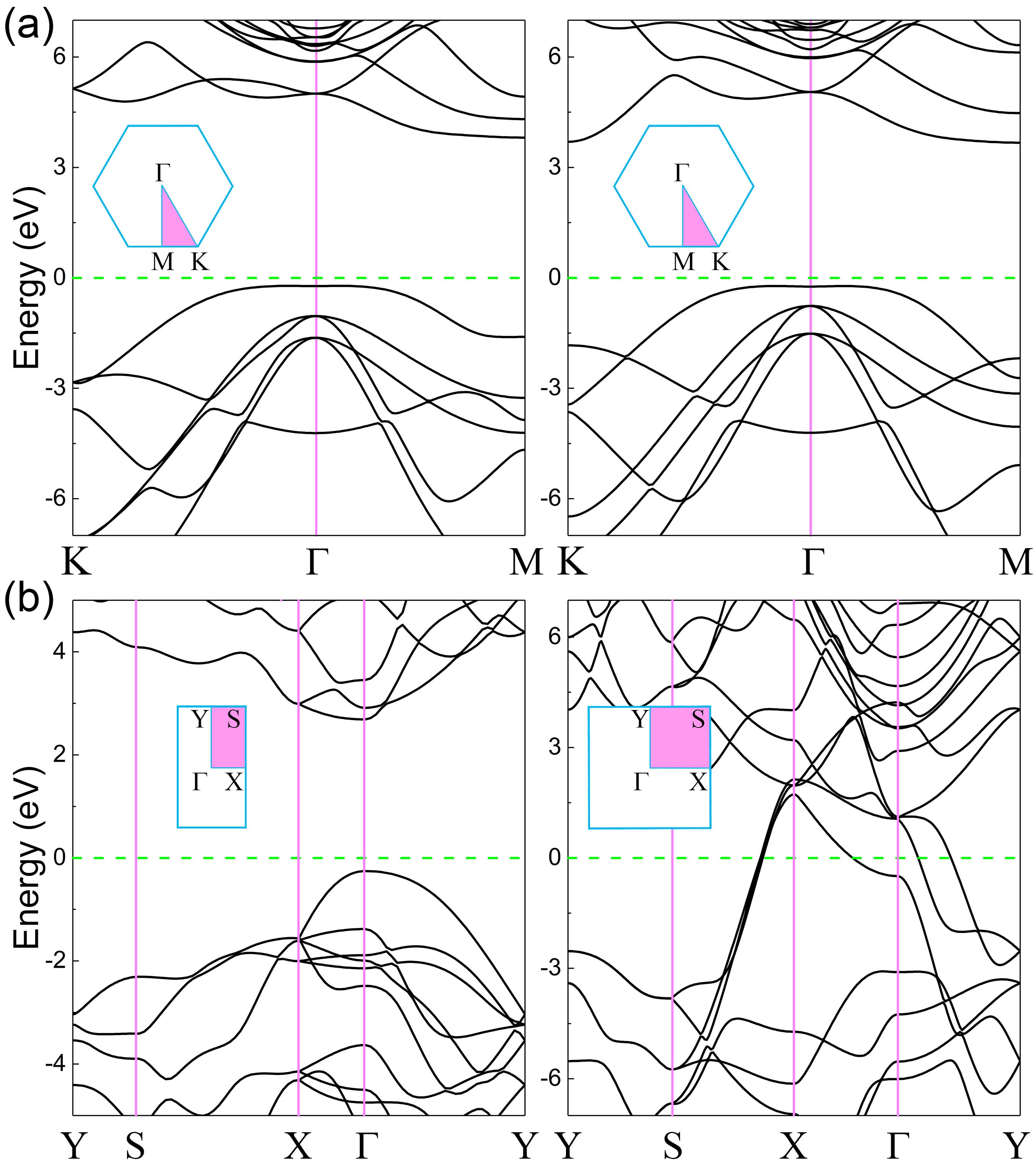}
\caption{Electronic band structures for (a) $\alpha$- (left), $\beta$- (right), (b) $\gamma$- (left), and $\delta$-BS (right) at HSE06 level.The Fermi energy level is set to zero. The high symmetry points in the first Brillouin-Zone of these phases are also drawn.}
\end{figure}

$\textbf{Electronic Structures.}$ The computed HSE06 band structures of the proposed sheets are shown in Figure 4. For $\alpha$- and $\beta$-BS, clearly, they are indirect band gap semiconductors: the conduction band minimum (CBM) is at the M point, while the valence band maximum (VBM) lies between the $\Gamma$ and K points, which is only slightly higher in energy than at the $\Gamma$ point (6.6 meV for $\alpha$-BS and 16.2 meV for $\beta$-BS at HSE06 level). However, for $\gamma$-BS, both CBM and VBM are located at the $\Gamma$ point, generating a direct band gap of 2.94 eV. For those three semiconductors, their band structures show strong anisotropy of the conduction band, which finally leads to the anisotropy of the effective masses. (see Figure S3) 

Interestingly, the top valence bands of $\alpha$- and $\beta$-BS sheets are nearly flat near around the $\Gamma$ point, leading to the Mexican-hat shape of the valence band edges, which render sharp peaks in the DOS and strong van Hove singularities near the Fermi level.\cite{miao2017tunable,cao2015tunable,kuc2017high} The Mexican-hat shape valence bands are mostly contributed by the 2p and 2p orbitals of B and S atoms, respectively, and these orbitals are strongly overlapping in the full energy range, suggesting covalent bonding characters of B-S bonds. (see Figure S4) The covalent features in the proposed 2D materials are also demonstrated by the analysis of the electron localization functions (ELFs), as shown in Figure S5. Obviously, ELFs show two localization areas: one is located around the B-S bonds and the other is between B-B bonds, reflecting the valence electrons are shared between the adjacent atoms.

$\textbf{Tunable Band Gaps and Superconducting.}$ One of the promising avenue to tune the electronic property of 2D materials is strain engineering. The band gaps of $\alpha$-, $\beta$-, and $\gamma$-BS structures with respect to the uniaxial stress are shown in Figure 5a. Approximately, the band gaps of the $\alpha$- and $\beta$-BS monolayers decrease gradually with either tensile or compressive strains, showing a nonmonotonic relationship. This unusual behavior is attributed to Mexican-hat shape valence bands near the Fermi level, akin to InP$_3$ monolayer.\cite{miao2017tunable} Their outstanding properties with heavy effective masses and wide band gaps render these materials suitable candidates for future applications in ultrashort ($\emph{i.e.}$, sub-5nm regime) channel logical devices.\cite{kuc2017high,fiori2014electronics}

\begin{figure}[htbp]
\centering
\includegraphics[width=0.9\columnwidth]{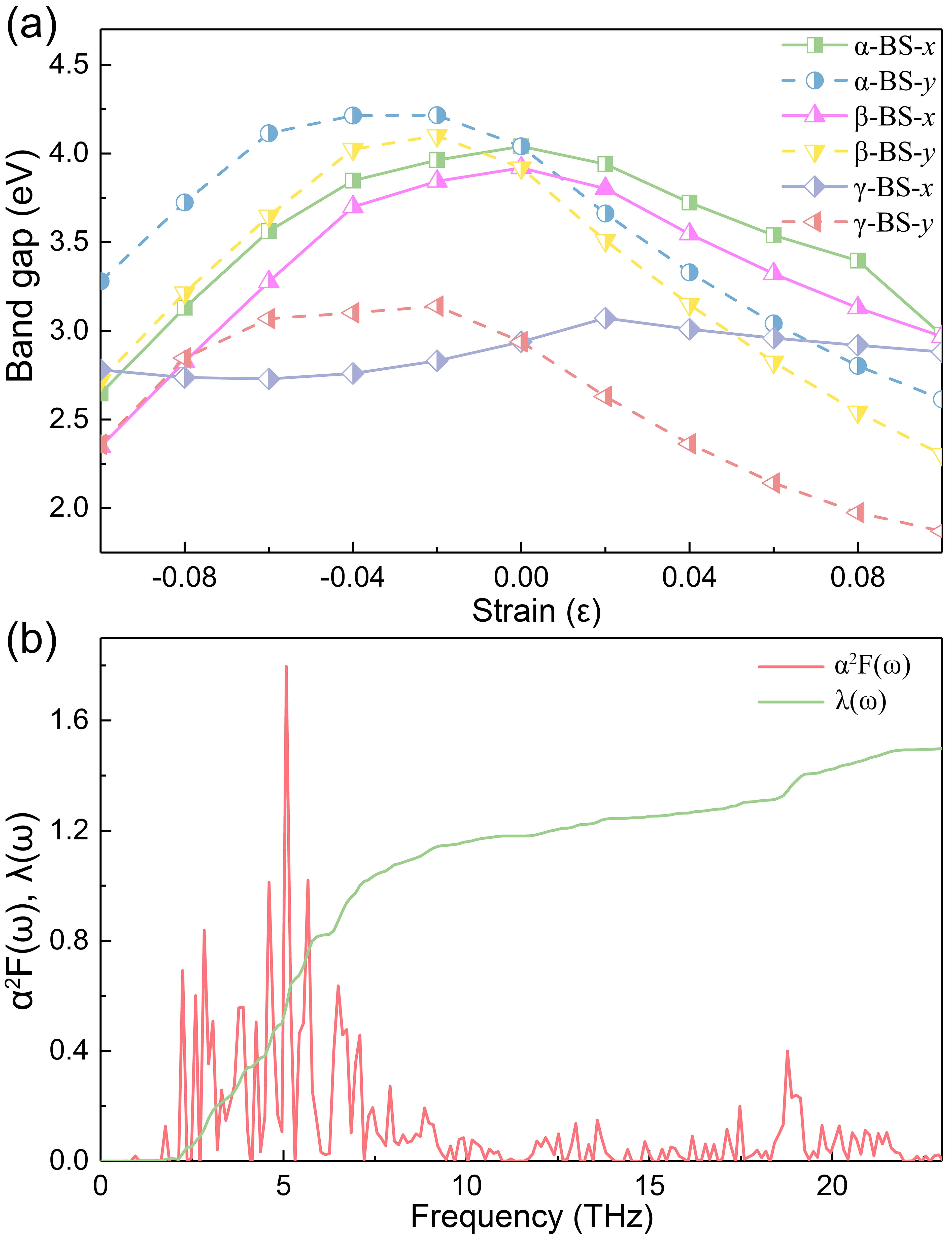}
\caption{(a) Dependence of the fundamental band gap on the in-plane uniaxial strain along the x (solid lines) and y (dashed lines) directions for $\alpha$-BS, $\beta$-BS, and $\gamma$-BS. (b) Calculated Eliashberg EPC spectral function and corresponding integral of the $\delta$-BS structure.}
\end{figure}

Additionally, the metallic property of $\delta$-BS inspire us to investigate its potential superconducting property. Figure 4b shows the Eliashberg spectral function $\alpha$$^2$$F$($\omega$) together with the integrated EPC parameter $\lambda$($\omega$) at the PBE level.  $\alpha$$^2$$F$($\omega$) exhibits a strong peak around 5 THz, and $\lambda$($\omega$) increases sharply in the range of 0-7 THz. As expected, the main contributor to the EPC is derived from the vibration of the heavy S atoms. The resulting coupling strength of $\lambda$ = 1.51, is rather strong. Superconducting transition temperature ($T_c$) of $\delta$-BS is estimated through the Allen-Dynes modified McMillan formula equation,\cite{allen1975transition} 
\begin{equation}
T_c=\frac{\omega{_{log}}}{1.2}exp(-\frac{1.04[1+\lambda]}{\lambda-\mu^{*}[1+0.62\lambda]})
\end{equation}
by using the calculated logarithmic average frequency ($\omega$$_{log}$) and a series of Coulomb pseudopotential parameters ($m*$) from 0.10 to 0.13, as shown in Table S2. At $m*$ = 0.10, the highest $T_c$ value of $\delta$-BS is 21.56 K, originating from its strong EPC and high logarithmic average frequency ($\omega$$_{log}$ = 189.06 K). Thus, the evaluated $T_c$ is in the range of 21.56 K ($m*$ = 0.10) to 19.08 K ($m*$ = 0.13), indicating that the $\delta$-BS is an intrinsic Bardeen-Cooper-Schrieffer (BCS) type superconductor. Notably, this $T_c$ is higher than that of other previously reported 2D superconductors, such as borophenes ($\approx $ 10-20 K), and boron carbides ($\approx $ 21.20 K).\cite{penev2016can,fan2018two}

\begin{figure}[htbp]
\centering
\includegraphics[width=0.9\columnwidth]{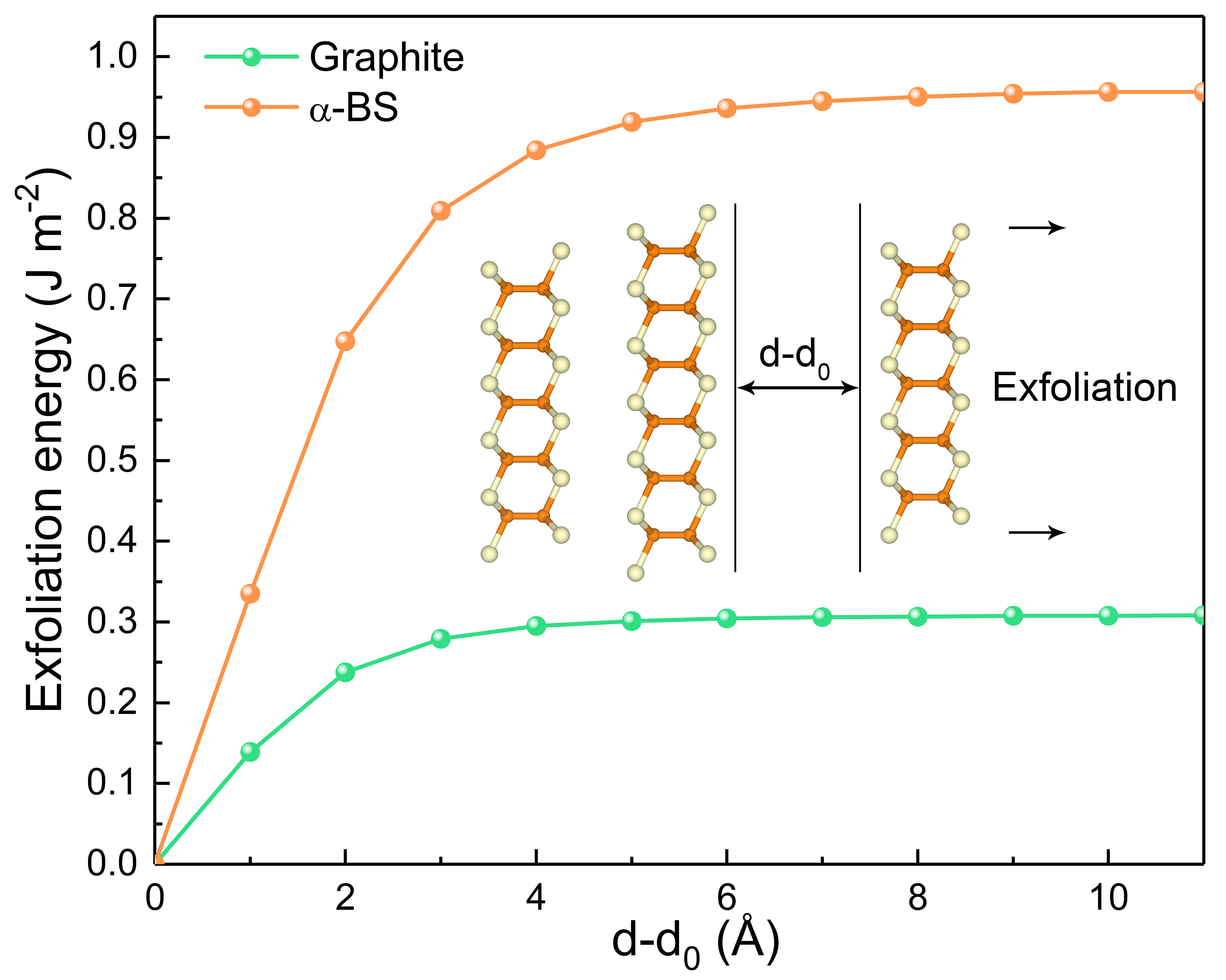}
\caption{Calculated exfoliation energy $\emph{vs}$ separation distance for $\alpha$-BS in comparison with graphite, where d$_0$ indicates the van der Waals (with vdw-DF correction\cite{dion2004van}) gap between adjacent layers in bulk $\emph{r}$-BS crystal.}
\end{figure}


$\textbf{Exfoliation of $\alpha$-BS.}$ The most common techniques to prepare 2D materials from their bulk counterparts are mechanical or liquid phase exfoliation.\cite{novoselov2004electric,kim2009large} Here, to explore the possibility of fabricating the energetically favorable $\alpha$-BS from the surface of its layered bulk $\emph{r}$-BS crystal (see Figure S6),\cite{sasaki2001high,cherednichenko2018boron} we then simulated the exfoliation process and calculated exfoliation energy with respect to separation, as shown in Figure 5. We first test the computing method using graphite as a benchmark, and the calculated exfoliation energy for graphene is 0.30 J m$^{-2}$, which is is consistent with the previous experimental (0.32 $\pm$ 0.03 J m$^{-2}$)\cite{zacharia2004interlayer} and theoretical value (0.31 J m$^{-2}$).\cite{miao2017tunable} For $\alpha$-BS, the calculated exfoliation energy is 0.96 J m$^{-2}$, which is higher than that of graphene, but still less than some layered materials, $\emph{i.e.}$, InP$_3$ (1.32 J m$^{-2}$),\cite{miao2017tunable} Ca$_2$N (1.08 J m$^{-2}$),\cite{zhao2014obtaining} and GeP$_3$ (1.14 J m$^{-2}$),\cite{jing2017gep3}  indicating the $\alpha$-BS sheet could be prepared experimentally from its bulk counterpart. Therefore, the moderate vdW interactions of $\alpha$-BS suggests the preparation of mono- or few-layer $\alpha$-BS heterostructures is feasible.\cite{tan2015epitaxial}

\section{CONCLUSION}

In summary, we have reported novel 2D BS binary sheets with high stability, high mechanical strength, and unique electronic properties. Importantly, $\delta$-BS phase is identified as the first discovery of intrinsic superconducting material among all 2D group \uppercase\expandafter{\romannumeral3A} chalcogenides. All monolayers show good dynamical and thermal stability, and $\alpha$-BS is expected to be prepared from its layered bulk $r$-BS by exfoliation. These advantaged features promote 2D BS sheets as promising candidates for future applications in future nano-devices. We also believe our results will further stimulate the experimentally preparation and investigation of 2D BS materials.

\section{ACKNOWLEDGMENTS}

This work was supported by the National Natural Science Foundation of China (Grant Nos. 11504325, 50972129, and 50602039), and Natural Science Foundation of Zhejiang Province (LQ15A040004). This work was also supported by the international science technology cooperation program of China (2014DFR51160), the National Key Research and Development Program of China (No. 2016YFE0133200), the European Union's Horizon 2020 Research and Innovation Staff Exchange (RISE) Scheme (No. 734578), and the One Belt and One Road International Cooperation Project from Key Research and Development Program of Zhejiang Province (2018C04021).

\bibliographystyle{unsrt}
\bibliography{ref}
\clearpage

\end{document}